# Discovery of ATLAS17jrp as an Optical-, X-Ray-, and Infrared-bright Tidal Disruption Event in a Star-forming Galaxy

Yibo Wang[1,2], Ning Jiang[1,2], Tinggui Wang[1,2], Jiazheng Zhu[1,2], Liming Dou[3], Zheyu Lin[1,2], Luming Sun[4], Hui Liu[5], and Zhenfeng Sheng[1,2]

[1] CAS Key laboratory for Research in Galaxies and Cosmology, Department of Astronomy, University of Science and Technology of China, Hefei, 230026, People's Republic of China; wybustc@mail.ustc.edu.cn, jnac@ustc.edu.cn, twang@ustc.edu.cn
[2] School of Astronomy and Space Sciences, University of Science and Technology of China, Hefei, 230026, People's Republic of China
[3] Department of Astronomy, Guangzhou University, Guangzhou 510006, People's Republic of China
[4] Department of Physics, Anhui Normal University, Wuhu, Anhui, 241002, People's Republic of China
[5] Anton Pannekoek Institute for Astronomy (API), University of Amsterdam, Science Park 904, 1098 XH, The Netherlands

## Abstract

We hereby report the discovery of ATLAS17jrp as an extraordinary tidal disruption event (TDE) in the star-forming galaxy SDSS J162034.99+240726.5 in our recent sample of mid-infrared outbursts in nearby galaxies. Its optical/UV light curves rise to a peak luminosity of $\sim 1.06 \times 10^{44}$ erg s$^{-1}$ in about a month and then decay as $t^{-5/3}$ with a roughly constant temperature around 19,000 K, and the optical spectra show a blue continuum and very broad Balmer lines with FWHM $\sim$ 15,000 km s$^{-1}$, which gradually narrowed to 1400 km s$^{-1}$ within 4 yr, all agreeing well with other optical TDEs. A delayed and rapidly rising X-ray flare with a peak luminosity of $\sim 1.27 \times 10^{43}$ erg s$^{-1}$ was detected $\sim$170 days after the optical peak. The high MIR luminosity of ATLAS17jrp ($\sim 2 \times 10^{43}$ erg s$^{-1}$) has revealed a distinctive dusty environment with a covering factor as high as $\sim$0.2, which is comparable to that of a torus in active galactic nuclei but at least one order of magnitude higher than normal optical TDEs. Therefore, ATLAS17jrp turns out to be one of the rare unambiguous TDEs found in star-forming galaxies, and its high dust-covering factor implies that dust extinction could play an important role in the absence of optical TDEs in star-forming galaxies.

*Unified Astronomy Thesaurus concepts:* Accretion (14); Galaxy nuclei (609)

*Supporting material:* data behind figures

## 1. Introduction

A tidal disruption event (TDE) occurs when a star occasionally comes too close to the supermassive black hole (SMBH) in the galaxy center. The star will be tidally disrupted, and nearly half of its material will be accreted onto the SMBH, producing a radiation flare peaked at the ultraviolet (UV) to soft X-ray band (e.g., Rees 1988; Gezari 2021). The rate of such occasions is about $\sim 10^{-4}$–$10^{-5}$ galaxy$^{-1}$ year$^{-1}$ (e.g., Wang & Merritt 2004; Stone & Metzger 2016). Although the first TDE was detected in the X-ray band (e.g., Bade et al. 1996), optical surveys have gradually dominated the discovery recently (e.g., Gezari 2021; van Velzen et al. 2021), benefiting from facilities devoted to sky surveys for transients, like the Zwicky Transient Facility (ZTF; Bellm et al. 2019). The optical light curves of TDEs usually show a monthly rise to a peak blackbody luminosity $L_{bb} \sim 10^{43.1}$–$10^{44.6}$ erg s$^{-1}$ (e.g., van Velzen et al. 2020, 2021), followed by a power-law decline as expected from such a fallback rate (e.g., Phinney 1989). The huge amount of energy released by TDEs illuminates the interstellar medium in the vicinity of the SMBH, providing an excellent chance to study the subparsec-scale nuclear environments, particularly for inactive galaxies.

If the SMBH is encircled by dust at the parsec scale, a significant part of the primary TDE radiation would be inevitably absorbed and reprocessed into the infrared (IR) band as echoes. The so-called IR echo was predicted to peak at the mid-IR band with a luminosity depending on the dust-covering factor ($f_c$) according to the 1D-transfer model (Lu et al. 2016). Immediately after the prediction, Jiang et al. (2016) achieved the first echo detection in ASASSN-14li with a peak luminosity of $2.5 \times 10^{41}$ erg s$^{-1}$ in IR utilizing archival Wide-field Infrared Survey Explorer (WISE; Wright et al. 2010; Mainzer et al. 2014) light curves. Subsequently, van Velzen et al. (2016) reported another two IR echoes of PTF09ge and PTF09axc with a low dust-covering factor of $f_c \sim 1\%$. Most recently, a systematic study of the IR echoes of 23 optical TDEs that happened in inactive galaxies (including LINERs) up to the end of 2018 has further established that their $f_c$ are at the order of 0.01 or even lower, in which the $f_c$ is simply defined as the ratio of the peak reprocessed IR to optical/UV blackbody luminosity (Jiang et al. 2021b). Obviously, the $f_c$ of known optical TDEs is much lower than that of a dusty torus in active galactic nuclei (AGNs, $f_c \sim 0.5$; e.g., Pier & Krolik 1993) and nuclei of normal quiescent galaxies like the Galactic ($f_c \sim 0.3$; e.g., Christopher et al. 2005). Hence, we may conclude that optical TDEs that occurred in inactive galaxies to date have been found only in dust-poor nuclear (parsec-scale) environments.

Besides their preference for dust-poor environments, TDEs are unexpectedly overrepresented in post-starburst galaxies with an enhancement factor of $\sim$100 but absent in the main sequence of star-forming (SF) galaxies (e.g., Arcavi et al. 2014; French et al. 2016, 2020; Hammerstein et al. 2021). Though some mechanisms, such as a high central stellar concentration or a circumnuclear gas disk surrounding the SMBH, have been proposed to interpret the TDE rate enhancement in post-starburst galaxies (see the review by French et al. 2020), they can hardly produce such a high enhancement factor or explain the nondetection of TDEs in normal SF galaxies. However, dust attenuation could play an important part in the absence of TDEs in SF galaxies, as evidenced by the discovery of obscured TDEs in ultraluminous IR galaxies (e.g., Tadhunter et al. 2017; Mattila et al. 2018) and the low dust-covering factor in the post-starburst galaxies as mentioned above. In addition, recent forward modeling by Roth et al. (2021) proposed that the dust extinction should be significant to suppress the TDE detection rate in SF galaxies. To directly examine this, robust TDE candidates in SF galaxies must be searched and found.

In this Letter, we report the discovery of an extraordinary TDE in the nearby SF galaxy SDSS J162034.99+240726.57 (J1620 +2407, hereafter) at redshift 0.06551. The event was initially noticed by the Asteroid Terrestrial Impact Last Alert System (ATLAS; Tonry et al. 2018; Smith et al. 2020) as ATLAS17jrp (Tonry et al. 2017), was suspected to be a TDE by Gromadzki et al. (2017), and was also selected into the sample of mid-infrared (MIR) outbursts in nearby galaxies (MIRONG; Jiang et al. 2021a; Wang et al. 2022). Its high luminosity in the optical, X-ray, and IR bands, as well as its SF host galaxy, makes it distinctive in comparison with all other optical TDEs, which might provide us with insightful clues to a complete understanding of TDE search and demography. We assume a cosmology with $H_0 = 70 \text{ km s}^{-1} \text{ Mpc}^{-1}$, $\Omega_m = 0.3$, and $\Omega_\Lambda = 0.7$.

## 2. Data

### 2.1. Optical, UV, and MIR Light Curves

We have collected multiband photometry spanning from UV to MIR, including data from ATLAS, Swift (Burrows et al. 2005; Roming et al. 2005), and WISE. All light curves are shown in Figure 1.

ATLAS17jrp was initially discovered by ATLAS on 2017 August 3; the discovery was announced publicly on the Transient Name Server (TNS) and given the name AT2017gge. We will refer to the transient by its survey name ATLAS17jrp throughout this paper. The ATLAS light curves were obtained using the ATLAS Forced Photometry Service, which produces point-spread function (PSF) photometry on either a target image or a difference image. The latter was selected to construct the ATLAS o- and c-band light curves from MJD 57,000 to MJD 58,900, and a median flux of the quiescent state was used to correct the zero-point. For epochs before MJD 57959 or after MJD58177.6, we binned the light curve every 15 days to improve the signal-to-noise ratio (S/N).

The ATLAS o-band light curve gives a good description of both the rising and declining phases. The flux at the o-band rose to the peak (MJD 57995, the epoch with maximum flux) within ∼36 days and then gradually faded to the quiescent state ∼200 days after the peak.

The Swift/UVOT photometry (2017–2018 PI: Blanchard; 2022 PI: Dou) was measured by the UVOTSOURCE task in the HEASoft package using 5″ apertures, and the AB magnitude was calibrated in the Swift photometric system. To improve the S/N, we do the photometry on the stacked image of every two epochs at late time. We use the latest observations as the reference to calculate the flux from the outburst.

We obtained the WISE light curves following our previous work (see Section 2.3 of Jiang et al. 2021a), and simply introduce the procedure as follows. The W1 and W2 profile-fit photometry of J1620+2407 was retrieved from the public AllWISE Multi-epoch Photometry Table and NEOWISE-R Single Exposure (L1b) Source Table, including all exposures from 2010 to 2021. The photometry was measured by PSF profile fitting, and the resulting light curves were binned every half year to improve the S/N after removing bad exposures with the quality flags. Then, an MIR light curve representing the outburst part was acquired by using the median magnitude of the quiescent state as a reference. The MIR light curve shows a peak at about MJD 58166 (i.e., the most luminous epoch) with a ∼171 day delay to the peak of the optical light curve. Furthermore, the peak MIR luminosity $L_{\rm dust,peak}$ is about $10^{43.32}$ erg s$^{-1}$ by applying a blackbody model to the W1 and W2 bands at peak.

### 2.2. X-Ray Light Curve and Spectral Analysis

We reprocessed the event files of Swift/XRT observations with the task xrtpipeline and selected those that operated in photon counting mode. The source was extracted with a region of radius of 20″, and the background was estimated in an annular region centered on the source position, with an inner radius of 80″ and an outer radius of 120″. No X-ray photon was detected in the first six observations (before MJD = 58143.7) or even on the stacked image with a total of 10.45 ks. Soft X-ray photons in the 0.3–0.8 keV band were detected in the 7th to 16th observations. In particular, 19 soft X-ray photons were detected in the ninth observation (MJD = 58168.659). However, photons harder than 0.8 keV were not detected in any observation. We calculated the net count rate and uncertainty of each observation with the Bayesian method, and adjacent observations were stacked together to improve the S/Ns (see the stacked region in Figure 1). The 7th–17th observations were stacked into an average spectrum with a count rate of $0.00452 \pm 0.00063$ cts s$^{-1}$ in a total exposure of 11.12 ks, and then we grouped the average spectrum to at least two counts per bin to adopt the C-statistic spectral fitting in XSPEC.

A simple Galactic absorbed power-law model with column density fixed at $4.90 \times 10^{20}$ cm$^{-2}$ (HI4PI Collaboration et al. 2016) was applied to the average spectrum. It resulted in a very steep photon index of $\Gamma = 5.5 \pm 0.7$ (Cstat/dof = 10.8/21), and an unabsorbed flux of $(5.0 \pm 0.7) \times 10^{-13}$ erg cm$^{-2}$ s$^{-1}$ in 0.3–2 keV band, corresponding to a luminosity of $6.4 \times 10^{42}$ erg s$^{-1}$ in 0.3–2 keV. We transformed the count rate of each observation into luminosity based on the power-law fitting result and show it in Figure 1. The unbinned X-ray light curve showed a large fluctuation after peak, which is probably due to the low exposure time for the postpeak observations (around 500 s except for the 11th and 12th ones). More recently (MJD = 59652), a 3.56 ks exposure of Swift/XRT gave an upper limit of $5.2 \times 10^{41}$ erg s$^{-1}$.

Replacing the power-law model with a blackbody resulted in a slightly better fit with a temperature of $kT = 70 \pm 8$ eV (Cstat/dof = 8.87/21). The total unabsorbed blackbody flux was $1.38 \times 10^{-12}$ erg cm$^{-2}$ s$^{-1}$, corresponding to a luminosity of $1.43 \times 10^{43}$ erg s$^{-1}$ for the average spectrum.

### 2.3. Optical Spectra Observation and Data Reduction

Four spectra were acquired for ATLAS17jrp, including one from SDSS DR7, one observed by ePESSTO groups with the ESO-NTT telescope,[6] and two from our own spectroscopic follow-up campaign of MIRONG (Wang et al. 2022). The ESO spectrum was observed at UT2017-09-14, near the peak of the optical light curve, while the two from the MIRONG project were

---

[6] We collected this spectrum from the TNS website: https://www.wis-tns.org/object/2017gge.

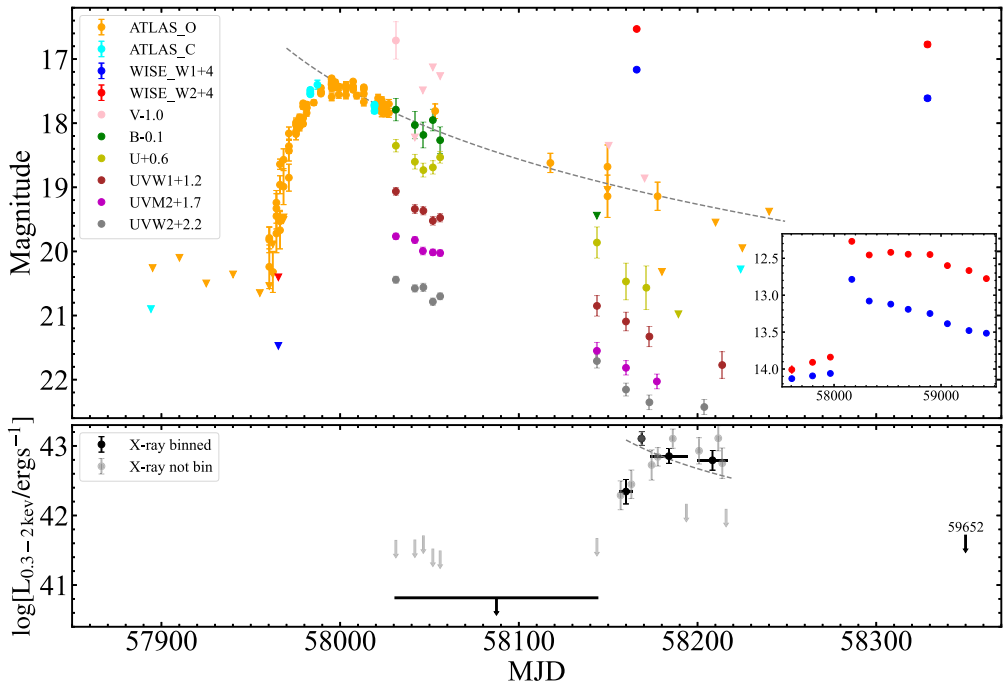

**Figure 1.** Top panel: host-galaxy-subtracted light curve of ATLAS17jrp in the optical, UV, and MIR. The triangles represent the upper limits (S/N of flux density less than 3), and the gray dashed curve shows the $t^{-5/3}$ power-law fit to the ATLAS o-band curve after the peak. The inset shows the long-term mid-infrared light curve of WISE without the host galaxy subtracted. Bottom panel: the Swift/XRT light curve of ATLAS17jrp. The gray points represent the detected 0.3–2 keV X-ray luminosity while the downward arrows indicate the upper limits. The binned X-ray light curve is shown as black points and arrows, with the bar indicating the binned region. The far-right arrow shows the upper limit from the recent Swift/XRT observation with its MJD marked nearby. The gray dashed curve represents a $t^{-5/3}$ power-law fitting to the binned X-ray light curve. The UV, optical, and MIR photometry and the Swift X-ray data are available as the data behind the figure.

(The data used to create this figure are available.)

observed at UT 2019-06-21 and UT 2021-03-17, respectively, both after the fading out of the UV/optical light curve. All spectra are shown in the top panel of Figure 2.

We analyzed the spectra according to a procedure similar to Wang et al. (2022), and we give a brief introduction here. We first did the Galactic extinction correction for these spectra with the recalibration version (Schlafly & Finkbeiner 2011) of the map of Schlegel et al. (1998) and the extinction curve of Fitzpatrick (1999). Because J1620+2407 is an SF galaxy, we fitted the continuum of the pre-outburst SDSS spectrum with the procedure PPXF (Cappellari & Emsellem 2004; Cappellari 2017) and the included MILES spectral library (Vazdekis et al. 2010). For the other spectra taken after the outburst, we modeled the continuum with a starlight component and a blackbody component. The starlight-component shape was fixed to that of the SDSS spectrum, and the blackbody temperature was fixed to the 19,000 K acquired from the blackbody fitting of the Swift data. Subtracting the continuum resulted in an emission-line spectrum, which was modeled by multiple Gaussian components, including narrow components (FWHM < 800 km s$^{-1}$) and broad components (FWHM > 1000 km s$^{-1}$). In the bottom panel of Figure 2, we showed the zoom-in of H$\alpha$ regions and the best-fit model applied to them.

## 3. Analysis

### 3.1. Host-galaxy Properties

The diagnostic narrow-line ratios can help separate SF galaxies from galaxies dominated by AGN activity, and the BPT diagram (Baldwin et al. 1981; Veilleux & Osterbrock 1987) with the narrow-line ratio from the pre-outburst SDSS spectrum classified J1620+2407 as an SF galaxy. Hence, there was no apparent AGN activity in the host galaxy J1620+2407 before the outburst. This is also supported by its nondetection in the radio band of FIRST. A dust-obscured AGN was further disfavored because J1620+2407 has a quiescent WISE color of W1 − W2 ∼ 0.16 (Stern et al. 2012; Yan et al. 2013). Additionally, though the optical TDEs preferentially occur in post-starburst galaxies, J1620+2407 is located in the region of normal SF galaxies (see Figure 3)

Acquired from Jiang et al. (2021a), the black hole mass of J1620+2407 is about $10^{6.67} M_\odot$, which was just well within the range preferred by TDEs (e.g., Gezari 2021).

### 3.2. ATLAS17jrp as a Robust TDE Candidate

First, the optical/UV light curves of TDEs usually show a monthly rise to a peak luminosity of about $10^{43.1\sim44.6}$ erg s$^{-1}$ and then decay with a timescale of about months to years, sometimes following a $t^{-5/3}$ power-law decline (e.g., Gezari 2021; van Velzen et al. 2021). Obviously, ATLAS17jrp showed similar rising and declining timescales and was even well consistent with a $t^{-5/3}$ power-law decline when such fitting was applied to the ATLAS o-band data (see Figure 1). Then, we did the blackbody fitting using the Swift data of six bands except for the V band, because it has a large uncertainty in the differential flux; the results are shown in Figure 4. The first Swift epoch showed a maximum blackbody luminosity $L_{bb} \sim 6.49 \times 10^{43}$ erg s$^{-1}$ and a blackbody radius $R_{bb} \sim 10^{15}$ cm, both followed by a decline. The blackbody temperature is about 19,000 ∼ 20,000 K and roughly remains constant for at least about a month. Undoubtedly, both the values of these blackbody parameters and their evolution behavior

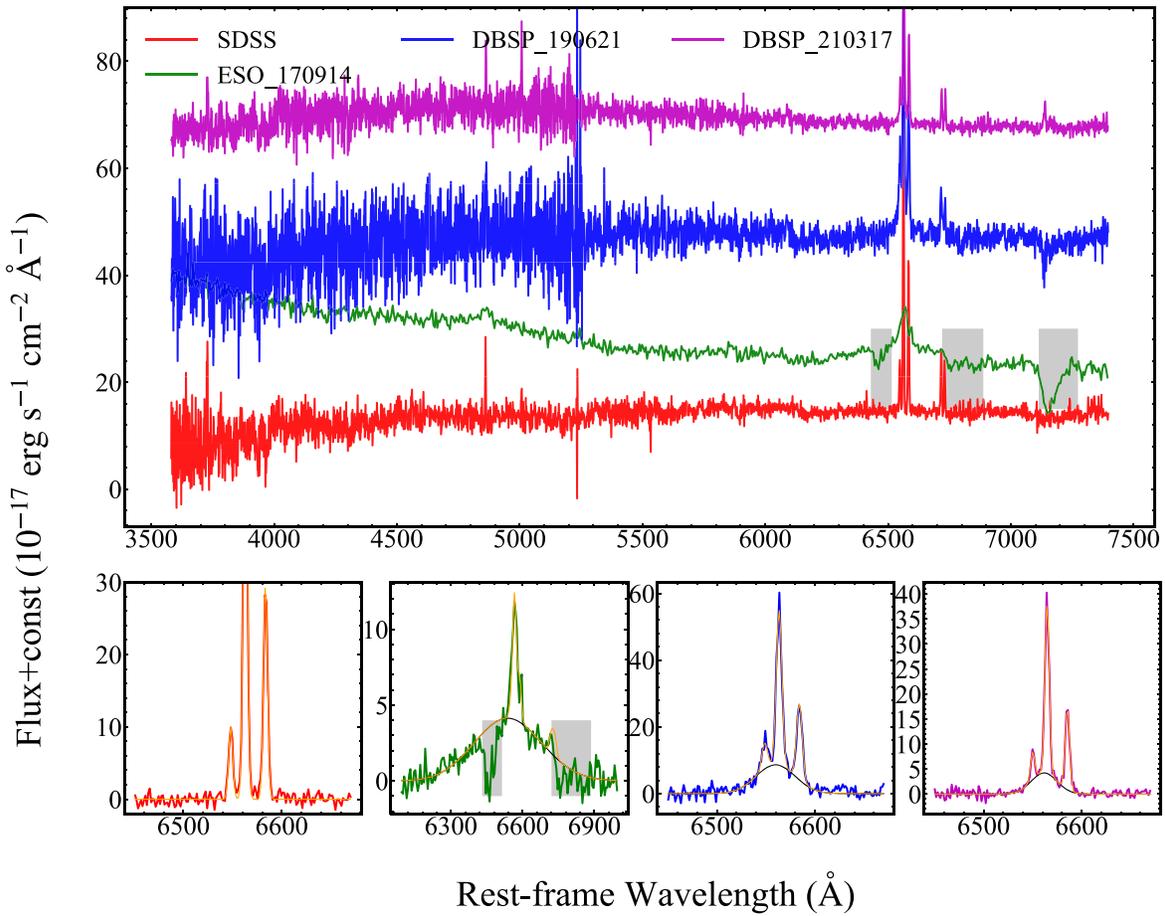

**Figure 2.** The spectra acquired for ATLAS17jrp. In the top panel, we showed the spectra without subtraction of host starlight. In the bottom panel, the Hα region was zoomed in and the Gaussian model fitting results were presented, with orange curves representing the best-fit model and black curves for broad components. The gray area covers the telluric absorption regions, which were masked in the fitting procedure. The two DBSP spectra are available as data behind the figure.

(The data used to create this figure are available.)

conform well to those of typical optical TDEs (e.g., van Velzen et al. 2020, 2021). Additionally, we estimated the peak luminosity to be $1.06 \times 10^{44}$ erg s$^{-1}$ from the first Swift epoch according to the $t^{-5/3}$ power-law decay fitting parameters.

On the other hand, the optical spectra of optical TDEs usually show a blue continuum and very broad Balmer or He II lines with FWHM ∼ 10,000 km s$^{-1}$ (e.g., van Velzen et al. 2020, 2021). For the source ATLAS17jrp, the ESO spectra taken near the peak of optical light curves, show a strong blue continuum, very broad Hα (FWHM ∼ 15,000 km s$^{-1}$), and broad Hβ (no reliable FWHM due to poor S/N), supporting the TDE scenario again. There are two other spectra taken at about 660 and 1295 days after the optical peak in the MIRONG project, when the light curves have already returned to the quiescent state. A narrower broad Hα component was detected in these two spectra with FWHM ∼ 2100 km s$^{-1}$ for the spectrum at 660 days and FWHM ∼ 1400 km s$^{-1}$ for the last one, and such a narrowing trend also favors the TDE scenario (e.g., van Velzen et al. 2020).

Additionally, a late-time X-ray brightening was found in ATLAS17jrp with about a 170 day delay between the X-ray and optical peak, which has also been seen before in some optical/UV TDEs, such as ASASSN-15oi (Gezari et al. 2017), AT2019azh (Liu et al. 2022), and OGLE16aaa (Kajava et al. 2020; Shu et al. 2020). As we have mentioned above, before MJD = 58143.708,

Swift/XRT had detected no X-ray photons but gave an upper limit of ∼$4.67 \times 10^{41}$ erg s$^{-1}$ in 0.3–2 keV in the sixth observation. Then, a peak luminosity of $1.27 \times 10^{43}$ erg s$^{-1}$ was reached with an enhancement factor of >27 in 25 days, indicating a rapid increase like OGLE16aaa but a lower peak luminosity like ASASSN-15oi and AT2019azh in 0.3–2 keV. Further, the detected X-ray was very soft, with a blackbody temperature of $kT \sim 70$ eV (or $\Gamma \sim 5.5$) for the average spectrum, which was similar to that of the X-ray components in both optical TDEs with late-time X-ray brightening and X-ray-selected TDEs. Such delayed brightening behavior of X-rays could be interpreted as a delayed accretion or obscuration of X-rays by optically thick clouds, which will be further discussed in our next paper.

To summarize, the optical/UV light curves, optical spectra, and X-ray properties of ATLAS17jrp show it was a robust TDE candidate.

### 4. Discussion

#### 4.1. High MIR Luminosity and Dust-covering Factor for ATLAS17jrp

Unlike the IR faintness of the previous optical TDEs, ATLAS17jrp shows an apparent MIR outburst with a peak luminosity $L_{\rm dust,peak} = 10^{43.32\pm0.1}$ erg s$^{-1}$, and actually, it is the most luminous one in the IR band for TDEs hosted by inactive (i.e., no AGN activity) galaxies until now. Before this work, the most IR-luminous optical TDE in a quiescent galaxy was the exceptional source AT2019dsg ($1.0 \times 10^{43}$ erg s$^{-1}$; van Velzen et al. 2021), which was found to be coincident with a neutrino event (Stein et al. 2021). But its high optical peak luminosity of $10^{44.54}$ erg s$^{-1}$ (Stein et al. 2021) indicates a dust-covering factor of just $\sim 0.03$, using the ratio of the peak IR luminosity $L_{dust,peak}$ to the optical peak luminosity $L_{opt,peak}$ as the authors did in Jiang et al. (2021b). However, using the same method, the dust-covering factor of J1620+2407 would be about 0.2. It is about one order of magnitude higher than that of previous optical TDEs including AT2019dsg (see Figure 5) and comparable to that of an AGN dust torus and that of the Galactic center. And also, the 171 day delay between the MIR and optical light curve limits these dust to subparsec scales near the SMBH. As we have mentioned above, the host of ATLAS17jrp is a normal SF galaxy rather than a post-starburst galaxy, as preferred by previous optical TDEs. Hence, the high dust-covering factor does suggest that dust obscuration/extinction plays an important role in the absence of TDEs in SF galaxies, namely that a significant part of TDEs in SF galaxies might be missed out in optical and X-ray surveys, which however could be revealed out by the MIRONG project (see below). Additionally, an echo model will be applied to the MIR light curve to acquire more detailed dust properties including dust geometry, in the next paper.

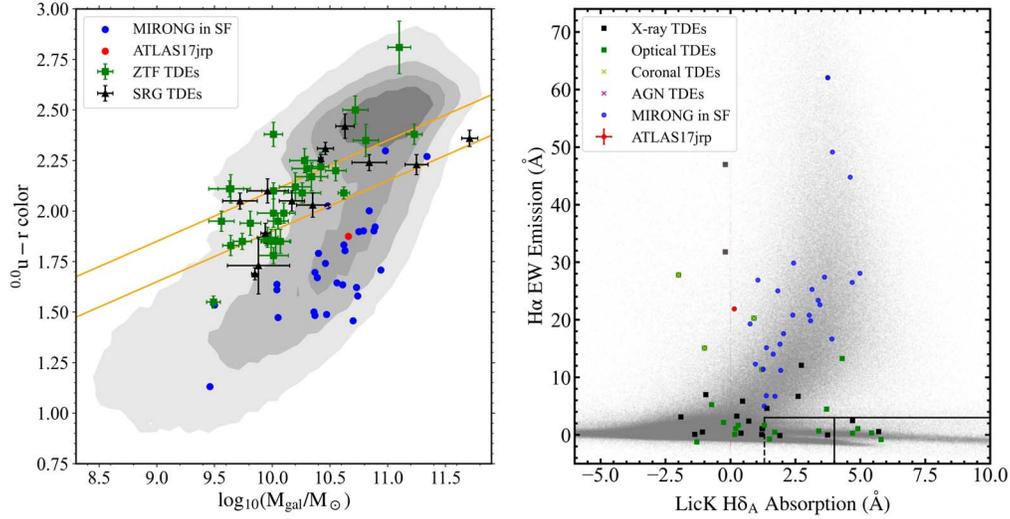

**Figure 3.** Left panel: the extinction-corrected, rest-frame $u - r$ color versus the total stellar mass of the TDE host galaxies. The green squares represent the 30 optical TDEs from the ZTF-I survey (Hammerstein et al. 2022) while the black triangles show the 13 X-ray TDEs found by the SRG/eROSITA (Sazonov et al. 2021). The orange line defines the green valley region according to van Velzen et al. (2021). Our MIRONG sample is shown in blue points except for ATLAS17jrp, which is highlighted in red. The MIRONG data points were acquired by matching with the SDSS+WISE galaxies catalog (Chang et al. 2015), which was also used as the comparison sample enclosed by the contours. Right panel: spectral indices of known TDE hosts (colored) and SDSS (gray) galaxies (French et al. 2016; Graur et al. 2018), i.e., H$\alpha$ EW (sensitive to current star formation) versus Lick H$\delta_A$ absorption (sensitive to star formation over the past gigayear) for each galaxy. We have shown the hosts of known TDE candidates (French et al. 2020; Sazonov et al. 2021) selected in the optical/X-ray bands with green/black filled squares. The hosts of TDEs in AGNs or selected by transient coronal-line emissions are marked by yellow and purple crosses. The blue points represent the MIRONGs (Jiang et al. 2021a; Wang et al. 2022) that occurred in SF galaxies, with ATLAS17jrp highlighted in red.

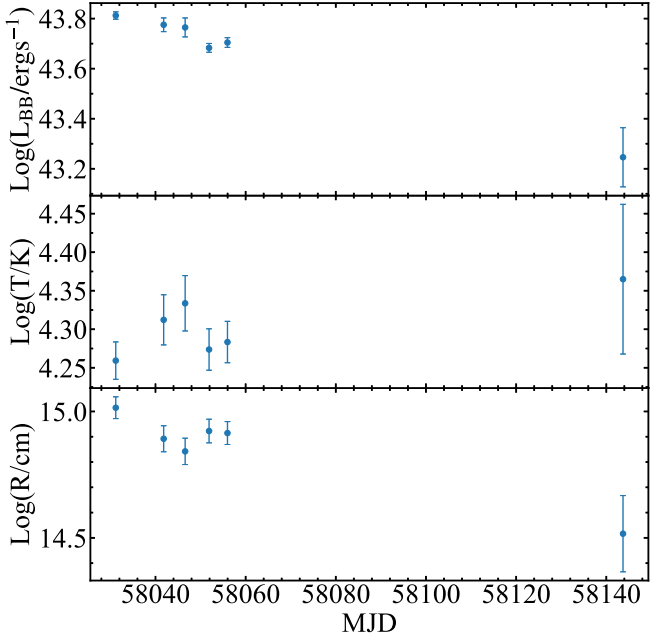

**Figure 4.** Blackbody-fitting parameters derived through fitting to the starlight-subtracted optical/UV fluxes of ATLAS17jrp.

### 4.2. Dusty TDEs in the MIRONG Sample

The event ATLAS17jrp was also selected in the MIRONG project, in which a large sample of MIRONG was constructed to search for TDEs or turn-on AGNs in dusty environments through the WISE database; subsequent photometry and spectroscopic follow-up are ongoing (see details in Jiang et al. 2021a; Wang et al. 2022). MIRONG usually occur in the galaxy center within a median offset of about 0″.1 and show a high MIR luminosity like J1620+2407 but higher than normal SNe. Therefore, the SN scenario was not favored for the majority of MIRONG. It was further precluded by the spectroscopic follow-up results because none of the SN spectral features like the P Cygni profile of Balmer lines or low-ionization metallic lines were found in the follow-up spectra (Wang et al. 2022). As a consequence, MIRONG in SF galaxies would be good TDE candidates with AGN variability excluded, and the discovery of ATLAS17jrp gave further evidence as an example because these MIRONG have similar peak MIR luminosity, integrated total IR energy, and dust properties to ATLAS17jrp (Jiang et al. 2021a). In Figure 3, we show the distribution of these MIRONG in SF galaxies without preference for post-starburst galaxies, and therefore, they could fill the absence of TDEs in SF galaxies, part of which were missed by optical or X-ray surveys due to dust/gas obscuration, like MIRONG. Further efforts would be made to search for more TDE candidates like ATLAS17jrp by combining the WISE and ZTF light curves and to make the physics behind MIRONG clear using multiband follow-up observations.

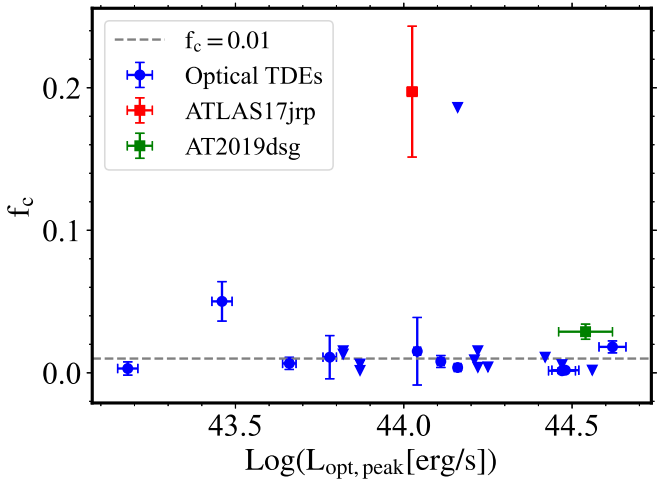

**Figure 5.** The peak optical luminosity vs. dust-covering factor for optical-bright TDEs. ATLAS17jrp stands out. The blue data points were taken from Jiang et al. (2021b), and downward triangles represent the 3σ upper limits.

*4.3. Implication of the Missing Energy of Optical/UV TDEs*

The observed total energy for optical/UV TDEs is only about $1 \times 10^{51}$ erg, which is one to two orders of magnitude lower than the predicted $10^{52-53}$ erg with the simple assumption that roughly about half of the debris was accreted onto the SMBH (e.g., Lu & Kumar 2018). Given this low apparent radiative efficiency, a scenario in which the optical emission is powered by stream–stream collision in the circularization process was suggested (e.g., Piran et al. 2015). Besides that, Lu & Kumar (2018) proposed that the missing energy could be radiated into unobserved EUV or in the form of a relativistic jet, and IR echoes could be an effective tool to examine this EUV scenario, which, however, is prevented by the bad IR light curves of previous optical/UV TDEs. Arp299-B AT1 (Mattila et al. 2018) and F01004 (Dou et al. 2017; Tadhunter et al. 2017), two TDE candidates in ultraluminous IR galaxies, did release energies of more than $10^{52}$ erg as indicated by their high IR integrated energy. However, the existence of AGN activity before the two outbursts makes the TDE nature less robust, and similarly, some other TDE candidates with a total energy of more than $10^{52}$ erg could not stand firm with their TDE classification. Nevertheless, ATLAS17jrp does have the typical optical/UV TDE features and good coverage of MIR light curves. Its latest integrated IR energy (until 2021 July 28) is about $1.3 \times 10^{51}$ erg. Considering the declining trend in IR light curves (see Figure 1) and the dust-covering factor of ATLAS17jrp, a total energy of more than $10^{52}$ erg would be expected, which is one order of magnitude higher than the observed total energy in optical/UV and X-ray ($\sim 10^{51}$ erg for optical and $\sim 10^{50}$ erg for X-ray), and therefore, the majority of the energy could be radiated as EUV. We will discuss this in more detail and investigate further, such as the constraints on the possible origin of the high unobserved energy, in the next paper.

## 5. Conclusion

In this work, we found an optical-, X-ray-, and infrared-bright TDE candidate, ATLAS17jrp, in a nearby SF galaxy J1620+2407. It possesses the highest peak MIR luminosity among known TDEs hosted by inactive galaxies with a dust-covering factor of ∼0.2, distinct from the typical value of ∼0.01 for previous optically selected TDEs in inactive galaxies. Hence, a significant fraction of TDEs in SF galaxies would be missed by optical surveys due to dust obscuration; they could, however, be revealed by the MIRONG project. The properties of ATLAS17jrp are concluded to be as follows:

1. The optical flux rises to a peak luminosity of $\sim 1.06 \times 10^{44}$ erg s$^{-1}$ in about a month and then decays as $t^{-5/3}$ with a radiation temperature of ∼19000 K.
2. The spectrum taken around the optical peak shows a strong blue continuum and very broad Balmer lines (FWHM ∼ 15000 km s$^{-1}$), which narrowed to 1400 km s$^{-1}$ in the last spectra.
3. ATLAS17jrp shows a delayed, rapidly rising X-ray flare with a peak luminosity of $1.27 \times 10^{43}$ erg s$^{-1}$ ∼170 days after the optical peak. The detected X-ray was very soft with a blackbody temperature of $kT \sim 70$ eV for the average spectrum.
4. ATLAS17jrp has a high peak MIR luminosity of $\sim 10^{43.32}$ erg s$^{-1}$ and could release a total energy of more than $10^{52}$ erg without missing energy, which is one order of magnitude higher than the observed energy in the optical and X-ray.

We thank the referee for helpful comments and suggestions, which led to the improvement of the paper. This work is supported by NSFC (11833007, 12073025, 12103048, 12192221), the B-type Strategic Priority Program of the Chinese Academy of Sciences (grant No. XDB41000000), China Manned Spaced Project (CMS-CSST-2021-B11), and the Fundamental Research Funds for the Central Universities (WK2030000023). This research uses data obtained through the Telescope Access Program (TAP). Observations with the Hale Telescope at Palomar Observatory were obtained as part of an agreement between the National Astronomical Observatories, Chinese Academy of Sciences, and the California Institute of Technology.

### ORCID iDs

Yibo Wang ● https://orcid.org/0000-0003-4225-5442
Ning Jiang ● https://orcid.org/0000-0002-7152-3621
Tinggui Wang ● https://orcid.org/0000-0002-1517-6792
Liming Dou ● https://orcid.org/0000-0002-4757-8622
Zheyu Lin ● https://orcid.org/0000-0003-4959-1625
Luming Sun ● https://orcid.org/0000-0002-7223-5840
Zhenfeng Sheng ● https://orcid.org/0000-0001-6938-8670